\documentclass[11pt,preprint]{aastex}

\begin{document}

\title{Gamma Ray Bursts and the Collapsar Model: \\
a comment on astro-ph/0203391 and astro-ph/0203467
}

\author{Charles D. Dermer}
\affil{E. O. Hulburt Center for Space Research, Code 7653,\\
Naval Research Laboratory, Washington, DC 20375-5352}
\email{dermer@osse.nrl.navy.mil}

\begin{abstract}

In an essay published in 1997, M.\ J.\ Harris pointed out that the development
of a theoretical explanation, or ``synthesis," of a group of astronomical
results can lead to a situation where subsequent observations are 
forced to fit within that framework. Moreover, new results 
may be reported that fit within the framework but which 
can no longer be reproduced following the
advent of more sensitive observations that overturn the synthesis.
Although Harris considered gamma-ray line transients, a similar situation
regarding the collapsar model for gamma-ray bursts now holds. In two recent papers posted 
on astro-ph by J.\ S.\ Bloom et al., and by P.\ A.\ Price et al., the synthesis
effect is illustrated with respect to the collapsar model.
These authors dismiss observations that
contradict the collapsar model, appeal to ``simplicity" to justify their preferred model,
and dismiss (or better yet, ignore) competing models with unscientific reasoning. Whether or not the collapsar model is correct, these papers serve as textbook examples of the synthesis
effect in science.
\end{abstract}

\newpage


\section{Introduction}

Gamma-ray burst (GRB) research is one of the most active fields in contemporary astronomy,
and overlaps a multitude of subjects in astrophysics, 
including compact object and black hole physics, star formation and cosmology,
particle acceleration and cosmic ray physics, and stellar evolution and supernovae. Fundamental progress
has been made in recent years to establish the distance scale to long-duration GRBs and the
radiation mechanisms that operate in relativistic blast waves. Less certainty holds with regard to 
the central engine, though it is generally accepted that GRB events involve the rapid
release of an enormous amount of energy within a small volume.

The establishment of the correct model for the central engine has acquired considerable urgency
following the discovery of X-ray afterglows with
the Beppo-SAX mission. The correct model will provide the solution to a 30-year old astronomical mystery, 
and will make sense of the wealth of multifrequency data now being collected on different
GRBs.

The scenario that presently receives the most consideration among GRB researchers, especially
GRB observers, is the collapsar model. This model involves the collapse of the central core of a massive,
evolved star to a newly-formed black hole. The black hole accretes matter at rates on the order of 
Solar masses per second to drive collimated, baryon-dilute outflows that penetrate the 
outer layers of the star to form a relativistic jet of plasma that, upon sweeping up material 
from the surrounding medium in an external shock, forms the afterglow emission.

Michael J.\ Harris wrote an article entitled ``Gamma-Ray Line Transients" in 1997 [1], where he surveyed the
early history of gamma-ray line searches, including annihilation line features. Prior to the 
launch of the {\it Compton Gamma Ray Observatory}, the various non-imaging balloon and space-borne spectrometers,
each with their own backgrounds and systematics, provided low-significance evidence for annihilation line transients.
Given the compelling explanation of variable features in terms of black-hole antimatter factories, a theoretical
synthesis was established that explained and motivated the detection of subsequent features, none of which
was confirmed by {\it CGRO}. Harris attributed this effect to ``the tendency to integrate imperfect information
into logical patterns." He cautioned scientists against this tendency, and to maintain ``(relative) objectivity, skepticism and especially our use of mathematical reasoning at every step in an argument."

This article came to mind after reading two papers recently posted on astro-ph. The first, entitled ``Detection 
of a Supernova Signature Associated with GRB 011121," by J.\ S.\ Bloom and 22 coauthors [2], and the second, entitled
``GRB 011121: A Massive Star Progenitor," by P.\ A.\ Price and 31 coauthors [3], present and interpret
 {\it HST} observations of the 
afterglow of GRB 011121. Although I can find nothing technically wrong with the observational procedures,
the authors of these papers seem to feel that they are obliged to explain their observations
within the context of the collapsar scenario. Once observers are in thrall to a particular model, their
objectivity is compromised. The point of this communication is to provide a cautionary note to observers to 
maintain a neutral position if their results are to be trusted. Science is not well served if 
results are forced into the service of models: the data should test the models, not the other way around.

Following a brief discussion of GRB source models in \S 2, taken from my recent review [4] and provided as background for
the general reader, I provide a commentary (\S 3) on statements made in these two articles. Let the reader judge
if the arguments made in these papers are scientifically sound. A summary is given in \S 4.

\section{Models for the GRB Central Engine}

In a recent review of GRB physics [4], I summarized and compared two models for the central engines of
GRBs, namely the collapsar model and the supranova model. Both are compatible with
evidence connecting GRBs to star-forming
regions [5,6] and, consequently, to a massive star origin. This argues against 
a third model involving coalescing compact objects, which should occur in both spiral and elliptical galaxies.
The absence of counterparts far from the disks of galaxy hosts also conflicts with this
model, insofar as old stellar systems can travel great distances before coalescence.
In particular, they should leave dusty regions where young and massive stars are born. 

A collapsar is a ``failed" supernova in the sense
that a core-collapse event fails to form a neutron star and instead produces a black hole. In Woosley's original paper [7],
the progenitor star was
suggested to be a rotating Wolf-Rayet star that produces, upon collapse, an accretion disk of 
several tenths of Solar masses. Hydrodynamical simulations of collapsars have specifically treated, for example, the evolution of a 35 $M_\odot$ main-sequence
star whose 14 $M_\odot$ helium core collapses to form a 2-3 $M_\odot$ black hole [8].  Provided that the core has a large amount of angular momentum, a delayed accretion event can be formed by the infalling matter. In order to produce a long-duration GRB with complex pulse structure under such circumstances, accretion
is argued to proceed over a period of time comparable to the prompt phase of a GRB.
The collapsar model must contend with the difficulty of ejecting baryon-clean material through an overlying shell of material. This is accomplished through an active central engine that persists for at least as long as the prompt phase of the GRB. Formation of relativistic jets of baryonic-clean material with bulk Lorentz factors  $\Gamma_0\sim 10^2$-$10^3$ represents a major difficulty in these models [9]. 

A central motivation of the supranova model of Vietri and Stella [10] is to identify a site that is originally free of baryon contamination. This occurs through a two-step collapse to a black hole, where a ``supramassive" neutron star (i.e., with mass exceeding several Solar masses) is formed in the first-step through a supernova explosion. The neutron star is initially  stabilized against collapse by rotation. The loss of angular momentum support through magnetic dipole and gravitational radiation leads to collapse to a black hole after some months to years. A two-step collapse process means that the neutron star is surrounded by a supernova shell of enriched material which can explain rebrightening events, as seen in GRB 970508 [11]. Alternately, the neutron star could be driven to collapse by accreting matter in a binary system [12]. The period of activity of a highly magnetized neutron star preceding its collapse to a black hole can produce a pulsar wind bubble consisting of a quasi-uniform low density, highly magnetized pair-enriched medium [13], in accord with afterglow model fits [14,15]. The earlier supernova could yield $\sim 1$ $ M_\odot$ of Fe in the surrounding vicinity. The discoveries of
variable Fe absorption in GRB 990705 during the prompt emission phase [16] and X-ray emission features in the afterglow spectra of GRB 991216 [17] provide some support for this model. 

\section{The Synthesis Effect}

The scientific process rests upon well-established methodological procedures
to determine whether a given model is falsifiable. In astronomy, where reproducibility
of an event may not be possible, a model should have a certain flexibility in terms of adjustable 
parameters to fit the data. Although the reasonableness of the fitted range of
the adjustable parameters introduces a subjective element 
into the assessment of a model, the quantitative evaluation of $\chi^2$ per degree of freedom, and the quality of
the improvement of the fit when introducing additional parameters is a well-established tool for observers
and data analysts.

Consider first the paper by Bloom et al.\ [2]. In the Introduction section of this paper,
circumstantial arguments for a massive star origin of long-duration GRBs are made. Only the collapsar model
is mentioned -- although the supranova model is also associated with a massive star origin, it is 
nowhere discussed in this section, leaving the reader to suppose that only the collapsar model is compatible with
these  observations. Both the collapsar and supranova models will have associated supernovae emissions,
so detection of a SN component is not sufficient to decide between the two models.
There are differences, however, between the observational signatures of the collapsar and supranova models. 
In the collapsar model, as the authors point out, 
an enriched circumburster medium due to stellar winds and mass loss is expected, which will show up 
in the afterglow spectra due to the $1/r^{2}$ dependence of the presupernova wind. The authors lament:
``{\it Unfortunately} broadband modeling of afterglows have not yielded firm signatures for such
circumburst medium." Unfortunately?  Why should it be unfortunate that observations have not
confirmed a theoretical prejudice? On the contrary, this is a tantalizing bit of information
about the environments in which GRB sources are located. 

The authors then describe bumps in afterglow spectra and various models to fit them.
One possible model is a dust echo, which seems entirely reasonable given that 
one-third to one-half of GRBs are dark bursts, that is, without optical afterglows.
These dark bursts may be caused by dust extinction. Rather than fit dust models
to test this option, which should have been entirely possible given the
range of personnel on the paper, the authors invoke a new method:
an appeal to ``simplicity"! In the authors' formulation, ``the simplicity of the supernova
model--requiring only a (physically motivated) adjustment in brightness--is a compelling
argument to accept our hypothesis." 

One might be willing to accept the validity of the supernova hypothesis for the late time
bump in the GRB light curves if the fits to the data were acceptable. But a cursory examination
of Figures 1 and 2 of the Bloom et al.\ paper show that the $\chi^2$ must be so great as to rule out the 
supernova hypothesis. No $\chi^2$ are in any case calculated, and the supernova hypothesis
seems to defy falsifiability: the authors invoke a renormalization of the luminosity. In the paper by Price et al.,
a new method is settled on to explain
discrepancies between the supernova hypothesis and observations: ``The absence of
SN components in other GRBs can be explained by appealing to the well known diversity in 
luminosity of Type 1b/c SNe" (no reference is given). In other words, any discrepancy between the model and observations
can be attributed to the wide range of light curves of Type Ib/c SNe, so they are not required to fit the data! 
Elsewhere in the paper by Bloom et al., it is asserted that 
"The consistency between the measurements 
and the SN [hypothesis] is {\it reasonable}, but some differences are observed, as expected." 
I do not find
the consistency reasonable, and I am not sure why differences are expected, given the
constant energy reservoir result [18]. Nor do the authors. Rather than contrast the collapsar and supranova
models in light of this result, the proffered explanation is that ``the constancy of the $\gamma$-ray
energy release is even more {\it mysterious}." 

Further evidence of the synthesis effect at work is found in the abstract and last section 
of the paper by Bloom et al., where they purport to ``exclude the (related) supranova model."
This is accomplished by a misrepresentation of the model, namely
that the supernova must take place ``months to years before the GRB event." Leaving aside
the fact that the duration between the first and second collapse events can range over arbitrary
time scales, the authors do not even calculate the range of allowed time delays between the hypothetical
supernova and GRB 011121. If they had, it might have occurred to them that different time delays
might explain the ``absence of SN components in other GRBs"  remarked upon by Price et al.\ (and which may
also be due to different slopes of the optical afterglows).

In the Price et al.\ paper [3], a new type of reasoning takes hold. Again they lament that
``Unfortunately, until now there has been no clear evidence for a wind-fed circumburst medium\dots"
But not to worry -- ``We undertake afterglow modeling of this important event and to our delight
have found a good case for a wind-fed circumburst medium." I would be delighted too, if the authors
had made a scientifically sound case. First, they use an analytic model to fit the data. The
analytic models, which lack light-travel time effects (among other things) found in numerical models [14],
 have been shown to be inaccurate in the radio regime by factors exceeding an order of magnitude [19]. The inaccuracy
is especially great near the endpoints and spectral breaks, precisely where the data must be  properly fit
for the authors to make their case that the explosion takes place in a wind-fed medium. Even granted the
uncertainty in the analytic fits, the wind model is favored only at the $2\sigma$ level, in other words,
the result is valid only at the 90\% confidence level. Perhaps
the glass is full at the 90\% confidence level; it might just as well be empty at the 10\% level.

In any case, this result does not square with the fact that many more GRBs that have been modeled and found
to be consistent with jetted emission in a uniform surrounding medium. As Price et al.\ argue, ``there could
be two different classes of progenitors within the class of long-duration GRBs (Chevalier and Li 2000 [15])."
Perhaps GRB 011121 is a collapsar, but then what about the more numerous second class?

\section{Summary}

The collapsar model has enjoyed a degree of acceptance that could lead one to conclude that the
correct model for the GRB engine has been found. 
 An effect of this situation is that observers either might discount their
 observations if they deviate from the model, or they might force the observations into the prevailing theoretical mold.
 I have argued that this ``synthesis effect" is already at work in the two papers 
discussed here. Without endorsing one model over another
(the author's views can be found in [4]), I would like to suggest to the GRB observers
that they either offer a more balanced appraisal of models, or report their data
free of theoretical biases.

\acknowledgments{The work of CD is supported by the Office of Naval Research.}

\newpage

{\bf REFERENCES}

\vskip0.2in

\noindent [1] Harris, M.\ J., 1997, in Proceedings of the Fourth Compton Symposium, ed.\ C.\ D.\ Dermer, M.\ S.\ Strickman, and J.\ D.\ Kurfess (AIP: New York), p.\ 418 

\noindent [2] Bloom, J.\ S., et al., 2002, Astrophysical Journal Letters, submitted (astro-ph/0203391, v.\ 1)

\noindent [3] Price, P.\ A. et al., 2002, Astrophysical Journal Letters, submitted (astro-ph/0203467, v.\ 1)

\noindent [4] Dermer, C.\ D., 2002, in Proc.\ 27th International Cosmic Ray Conference, Hamburg, Germany (7-15 August 2001)(astro-ph/0202254) 

\noindent [5] Djorgovski, S.\ G., et al.\ 2001, in Gamma Ray Bursts in the Afterglow Era, ed. E. Costa, F. Frontera, and J. Hjorth (Springer: Berlin), 218

\noindent [6] Sokolov, V.~V.~et al.\ 2001, Astronomy and Astrophysics, {\bf 372}, 438

\noindent [7] Woosley, S.E. 1993, Astrophysical Journal, {\bf 405}, 273

\noindent [8] MacFadyen, A.~I.~and Woosley, S.~E., 1999, Astrophysical Journal, {\bf 524}, 262

\noindent [9] Tan, J.~C., Matzner, C.~D., and McKee, C.~F., 2001, Astrophysical Journal, {\bf 551}, 946

\noindent [10] Vietri, M., and Stella, L., 1998, Astrophysical Journal, {\bf 507}, L45 

\noindent [11] Vietri, M., Perola, C., Piro, L., and Stella, L., 1999, Monthly Notices of the Royal Astronomical Society, {\bf 308}, L29

\noindent [12] Vietri, M.~and Stella,  L., 1999, Astrophysical Journal, {\bf 527}, L43 

\noindent [13] K\"onigl, A., and Granot, J., 2001, Astrophysical Journal, submitted (astro-ph/0112087)

\noindent [14] Panaitescu, A., and Kumar, P., 2001, Astrophysical Journal, {\bf 554}, 667

\noindent [15] Chevalier, R., and Li, Z.-Y., 2000, Astrophysical Journal, {\bf 536}, 196

\noindent [16] Amati, L., et al., 2000, Science, {\bf 290}, 953

\noindent [17] Piro, L., et al., 2000, Science, {\bf 290}, 955

\noindent [18] Frail, D.\ A., et al., 2001, Astrophysical Journal, {\bf 562}, L55

\noindent [19] Dermer, C. D., B\"ottcher, M., and Chiang, J., 2000, Astrophysical Journal, {\bf 537}, 255

\end{document}